\RequirePackage{lineno}
\documentclass[twocolumn,aps,tightenlinefs,superscriptaddress,preprintnumbers]{revtex4}
\usepackage{graphicx,color,dcolumn,booktabs,bm}
\usepackage{longtable,lscape}
\usepackage{amsmath}
\usepackage{indentfirst}
\usepackage{epsfig}
\usepackage{feynmf}
\usepackage{epstopdf}
\usepackage{slashed}
\usepackage{cases}
\usepackage{color}
\usepackage{multirow}
\usepackage{ulem}
\usepackage{graphicx,color,dcolumn,booktabs,bm}
\usepackage{verbatim}
\usepackage[colorlinks,linkcolor=blue,anchorcolor=green,citecolor=blue]{hyperref}
\usepackage{mathrsfs}
\usepackage{rotating}
\usepackage{threeparttable}
\usepackage{dcolumn}
\usepackage{orcidlink}
\usepackage{graphicx}
\usepackage{overpic}
\usepackage{hyperref}

\newcommand{\beq}{\begin{equation}}
\newcommand{\eeq}{\end{equation}}
\newcommand{\bitm}{\begin{itemize}}
\newcommand{\eitm}{\end{itemize}}

\begin{document}
\hyphenpenalty=10000


\title{\quad\\[0.1cm]\boldmath First observation of the decay
$D_{s1}(2460)^{+} \to D_{s}^{*+}\gamma$ and measurement of the branching-fraction ratio $\mathcal{B}[D_{s1}(2460)^{+} \to D_{s}^{*+} \gamma]/\mathcal{B}[D_{s1}(2460)^{+} \to D_{s}^{+} \gamma]$}

  \author{I.~Adachi\,\orcidlink{0000-0003-2287-0173}} 
  \author{A.~Aggarwal\,\orcidlink{0000-0002-5623-3896}} 
  \author{H.~Ahmed\,\orcidlink{0000-0003-3976-7498}} 
  \author{Y.~Ahn\,\orcidlink{0000-0001-6820-0576}} 
  \author{H.~Aihara\,\orcidlink{0000-0002-1907-5964}} 
  \author{M.~Akdag\,\orcidlink{0009-0004-3728-1077}} 
  \author{N.~Akopov\,\orcidlink{0000-0002-4425-2096}} 
  \author{A.~Akram\,\orcidlink{0000-0003-0198-5852}} 
  \author{S.~Alghamdi\,\orcidlink{0000-0001-7609-112X}} 
  \author{M.~Alhakami\,\orcidlink{0000-0002-2234-8628}} 
% ...
\collaboration{The Belle and Belle II Collaborations}

\begin{abstract}
We report the first observation of the radiative decay $D_{s1}(2460)^{+} \to D_{s}^{*+} \gamma$ with a significance greater than 10 standard deviations. The signal is found in $e^{+}e^{-} \to c\bar{c}$ production using data samples with integrated luminosities of 980 and 572~$\mathrm{fb}^{-1}$ collected by the Belle and Belle~II detectors operating at the KEKB and SuperKEKB asymmetric-energy $e^{+}e^{-}$ colliders, respectively. The branching-fraction ratio $\mathcal{B}[D_{s1}(2460)^{+} \to D_{s}^{*+} \gamma]/\mathcal{B}[D_{s1}(2460)^{+} \to D_{s}^{+} \gamma]$ is measured to be $0.74 \pm 0.05 ({\rm stat.}) \pm 0.06 ({\rm syst.)}$. This result helps discriminate between different theoretical models regarding the internal structure of the charm-strange meson $D_{s1}(2460)^{+}$.
\end{abstract}

\maketitle
Exotic hadrons provide sensitive probes of the nonperturbative dynamics of quantum chromodynamics (QCD), 
shedding light on the formation of color-singlet matter beyond the conventional quark model\mbox{~\cite{Olsen:2017bmm, Guo:2017jvc, Brambilla:2019esw, Chen:2022asf, Dai:2026fkg}}.
The discoveries of the charm-strange mesons $D_{s0}^{*}(2317)^{+}$~\cite{BaBar:2003oey}  and $D_{s1}(2460)^{+}$~\cite{CLEO:2003ggt} have 
long challenged conventional quark-model expectations, as their masses are lower than the predictions for the 
corresponding $c\bar{s}$ states with matching spin ($J$) and parity ($P$)~\cite{Godfrey:1985xj, Godfrey:1986wj, DiPierro:2001dwf, Ebert:2009ua}. 
Both states are very narrow and lie below the $DK$ and $D^{*}K$ thresholds, respectively, 
with their dominant decays proceeding through isospin-violating processes~\cite{PDG}.
These properties have motivated various interpretations of their nature,  including
$c\bar{s}$ mesons~\cite{Godfrey:2003kg, Bardeen:2003kt, Ferrandes:2004aa, Green:2016occ, Radford:2009bs, Fayyazuddin:2003aa, Song:2015nia, Wei:2005ag, Godfrey:2015dva, Chen:2020jku, Bondar:2025qzg, Colangelo:2005hv}, hadronic molecules~\cite{Close:2005se,Barnes:2003dj, Kolomeitsev:2003ac, Guo:2006fu, Zhang:2006ix, Rosner:2006vc, Faessler:2007us, Fu:2021wde, Zhang:2024usz}, compact tetraquark configurations~\cite{Cheng:2003kg, Chen:2004dy, Kim:2005gt, Chen:2026wrh}, and mixed states~\cite{MartinezTorres:2014kpc, Song:2022yvz,Yang:2023tvc,Gil-Dominguez:2023puj,Wang:2025jcq, Albaladejo:2018mhb, Yang:2021tvc}.  
Even after extensive experimental and theoretical efforts~\cite{Godfrey:2003kg, Bardeen:2003kt, Ferrandes:2004aa, Green:2016occ, Radford:2009bs, Fayyazuddin:2003aa, Song:2015nia, Wei:2005ag, Godfrey:2015dva, Chen:2020jku, Bondar:2025qzg, Colangelo:2005hv, Barnes:2003dj, Kolomeitsev:2003ac, Guo:2006fu, Zhang:2006ix, Rosner:2006vc, Faessler:2007us, Fu:2021wde, Fu:2025lfo, Guo:2026ngz, Aliev:2026zyc, Zhang:2024usz, Cheng:2003kg, Chen:2004dy, Kim:2005gt, Chen:2026wrh, MartinezTorres:2014kpc,  Song:2022yvz,Yang:2023tvc,Gil-Dominguez:2023puj,Wang:2025jcq, Albaladejo:2018mhb, Yang:2021tvc, Close:2005se, Lutz:2007sk, Belle:2003guh, BaBar:2004yux, BaBar:2006jvx, Belle:2003kup, BaBar:2006eep, BESIII:2017vdm, Belle II:Ds2317, LHCb:2024iuo}, 
the internal structure of the $D_{s0}^{*}(2317)^{+}$ and $D_{s1}(2460)^{+}$ mesons remains an open question in charm spectroscopy. 

Radiative decays play a distinctive role in elucidating hadron structure, as their electromagnetic transition rates are sensitive to the underlying internal configuration~\cite{Fu:2025lfo, Guo:2026ngz, Aliev:2026zyc}.
The radiative decay $D_{s0}^{*}(2317)^{+} \to D_{s}^{*+} \gamma$  was recently observed for the first time by the Belle and Belle~II experiments, 
providing important constraints on theoretical interpretations of the $D_{s0}^{*}(2317)^{+}$ structure~\cite{Belle II:Ds2317}.
The $D_{s1}(2460)^+$, which has $J^P=1^+$~\cite{Belle:2003guh,BaBar:2004yux}, can undergo radiative decays to both $D_s^+\gamma$ and $D_s^{*+}\gamma$.
The decay $D_{s1}(2460)^{+} \to D_s^{+}\gamma$ was first observed by Belle~\cite{Belle:2003guh}. 
Searches for $D_{s1}(2460)^{+} \to D_s^{*+}\gamma$ were performed by CLEO~\cite{CLEO:2003ggt}, Belle~\cite{Belle:2003kup}, and 
BaBar~\cite{BaBar:2006eep}, but no significant signal was observed.  The branching-fraction ratio $\mathcal{R} = \mathcal{B}[D_{s1}(2460)^{+} \to D_{s}^{*+} \gamma]/\mathcal{B}[D_{s1}(2460)^{+} \to D_{s}^{+} \gamma]$ 
is expected to be sensitive to the internal structure of the $D_{s1}(2460)^{+}$. Existing theoretical calculations provide
a wide range of predictions for $\mathcal{R}$\mbox{~\cite{Godfrey:2003kg, Bardeen:2003kt, Radford:2009bs, Ferrandes:2004aa, Green:2016occ, Bondar:2025qzg,
Fu:2021wde, Colangelo:2005hv, Close:2005se, Zhang:2024usz, Lutz:2007sk}}.
Most calculations within the conventional $c\bar{s}$ picture yield values of $\mathcal{R}$ between approximately 0.35 and 1.33~\cite{Godfrey:2003kg, Bardeen:2003kt, Radford:2009bs, Ferrandes:2004aa, Green:2016occ, Bondar:2025qzg}, whereas some calculations based on the $D^{*}K$ molecular interpretation predict smaller values, typically around 0.07--0.31~\cite{Fu:2021wde, Zhang:2024usz}. A measurement of $\mathcal{R}$ can provide constraints on the internal structure of the $D_{s1}(2460)^{+}$ and test competing interpretations of its nature.

The interpretation of the $D_{s1}(2460)^{+}$ within the conventional quark-model framework suggests that it can be a mixture of $^{1}\mathrm{P}_{1}$ and
$^{3}\mathrm{P}_{1}$ $c\bar{s}$ configurations\mbox{~\cite{Close:2005se, Bondar:2025qzg}}.
Since the radiative decays \mbox{$D_{s1}(2460)^{+}\to D_s^{*+}\gamma$ and}
$D_{s1}(2460)^{+}\to D_s^{+}\gamma$ couple differently to these two components, the value of $\mathcal{R}$ depends on the underlying mixing angle. For instance, Close and Swanson~\cite{Close:2005se} calculate $\Gamma(D_{s1}(2460)^{+}\to D_s^{*+}\gamma)=4.2\sin^{2}\theta_s~\mathrm{keV}$ and $\Gamma(D_{s1}(2460)^{+}\to D_s^{+}\gamma)=7.3\cos^{2}\theta_s~\mathrm{keV}$,
giving $\mathcal{R}=0.58\tan^{2}\theta_s$, where $\theta_s$ is the
${}^{1}\mathrm{P}_{1}$--${}^{3}\mathrm{P}_{1}$ mixing angle.
Thus, a measurement of $\mathcal{R}$ would help constrain the possible 
${}^{1}\mathrm{P}_{1}$--${}^{3}\mathrm{P}_{1}$ mixing pattern of $D_{s1}(2460)^{+}$.

In this Letter, we report the first observation of the radiative decay $D_{s1}(2460)^{+} \to D_{s}^{*+} \gamma$, with $D_{s}^{*+} \to D_{s}^{+} \gamma$, in continuum $e^{+}e^{-} \to c\bar{c}$ production. The decay $D_{s1}(2460)^{+} \to D_{s}^{+} \gamma$ is used as the reference channel, 
and the ratio of the two decay branching fractions, $\mathcal{R}$,
is measured. The $D_{s}^{+}$ candidates are reconstructed via the $\phi \pi^{+}$ and $K^{+} \overline{K}^{*0}$ decay channels, both of which lead to
the same $K^{+}K^{-}\pi^{+}$ final state. Charge-conjugate decays are implicitly included throughout this Letter. This measurement is based on data recorded with the Belle detector at center-of-mass (c.m.) energies ranging from $9.46$ to $10.87~\mathrm{GeV}$~\cite{Belle:2000cnh, Belle:2012iwr}, and with the Belle~II detector from $10.52$ to $10.81~\mathrm{GeV}$~\cite{Belle-II:2010dht}. The integrated luminosities are 980 and 572$~\mathrm{fb}^{-1}$ for Belle and Belle~II, respectively.


Data and Monte Carlo (MC)  simulated  samples for both Belle and Belle~II are processed with the Belle~II analysis software framework~\cite{Kuhr:2018lps, Gelb:2018agf}. 
MC simulated signal events are used to optimize the selection criteria, calculate the detection efficiencies, and determine the fit models. 
Continuum $e^{+}e^{-} \to c\bar{c}$ events are generated using \textsc{pythia6}~\cite{Sjostrand:2006za} for Belle and \textsc{pythia8}~\cite{Sjostrand:2014zea} and \textsc{kkmc}~\cite{Jadach:1999vf}
for Belle~II, where at least one of the charm quarks hadronizes into a $D_{s1}(2460)^{+}$ meson for the signal and reference channels. 
The decay $D_{s1}(2460)^{+} \to D_{s}^{*+}\gamma$ is generated with the phase space model, while the decays $D_{s}^{*+}\to D_{s}^{+}\gamma$ and $D_{s1}(2460)^{+} \to D_{s}^{+}\gamma$ are generated as P-wave decays. The decay $D_{s}^{+} \to K^{+} K^{-} \pi^{+}$ is modeled based on the previous measurement~\cite{BESIII:2020ctr}.
The effect of final-state radiation is taken into account in the simulation using the \textsc{photos} package~\cite{Barberio:1990ms}. Generated signal events undergo detector simulation with \textsc{geant3}~\cite{GEANT3} 
for Belle and \textsc{geant4}~\cite{GEANT4:2002zbu} for Belle~II.  The signal MC events are reweighted according to the $x_{p}$ distribution measured in the reference channel, where $x_{p} \equiv p^{*}_{D_{s1}(2460)^{+}}/p^{*}_{\rm max}$ is the reduced momentum of the selected $D_{s1}(2460)^{+}$ candidate. Here, $p^{*}_{D_{s1}(2460)^{+}}$ is its momentum in the $e^{+}e^{-}$ c.m.\ frame,  and $p^{*}_{\rm max} \equiv \frac{1}{c}\sqrt{E_{\rm beam}^{2}-M^{2}[D_{s1}(2460)^{+}]c^{4}}$, where $E_{\rm beam}$ is the beam energy in the $e^{+}e^{-}$ c.m.\ frame and $M[D_{s1}(2460)^{+}]$ is the invariant mass of the $D_{s1}(2460)^{+}$ candidate. 
To study background sources~\cite{Zhou:2020ksj} and optimize the selection criteria,
inclusive MC samples as described in Ref.~\cite{Belle:2024xcs} are used.

To avoid experimental bias, the signal region in data is not examined until the analysis procedure is finalized. All selection criteria are determined through an iterative optimization of a figure of merit~\cite{Punzi} for a target significance of five standard deviations ($\sigma$) for the signal decay $D_{s1}(2460)^{+}\to D_s^{*+}\gamma$, based on simulated samples. We apply identical selection criteria in the Belle and Belle~II analyses.

The $D_s^{+}$ candidates are reconstructed from the $K^{+}K^{-}\pi^{+}$ final state using the same selection criteria as those used in the recent observation of the radiative decay $D_{s0}^{*}(2317)^{+} \to D_{s}^{*+}\gamma$~\cite{Belle II:Ds2317}.
The candidates are required to be consistent with either the $\phi\pi^{+}$ or $K^{+}\overline{K}^{*0}$ decay channel. For the $K^{+} \overline{K}^{*0}$ channel, an additional requirement of $|\cos\theta_{h}(\overline{K}^{*0})|>0.3$ is imposed to suppress combinatorial background.
Here, $\theta_{h}(\overline{K}^{*0})$ is the helicity angle of the $\overline{K}^{*0}$ candidate, i.e., the angle between the $K^{-}$ momentum and the direction opposite to the $D_s^{+}$ momentum in the $\overline{K}^{*0}$ rest frame. The signal region for the reconstructed $D_{s}^{+}$ candidates is defined as
$|M^{\rm rec}(K^{+}K^{-}\pi^{+})-m(D_{s}^{+})|< 10~{\rm MeV}/c^2$, corresponding to approximately 2.5 times the mass resolution. Here and throughout the following, $M^{\rm rec}(X)$ denotes the reconstructed invariant mass of the system $X$.
The sidebands are defined as $|M^{\rm rec}(K^{+}K^{-}\pi^{+}) - m(D_{s}^{+}) \pm 40| < 10~{\rm MeV}/c^2$, with a total width twice that of the signal region. Here and throughout this paper, $m(i)$ denotes the known mass of particle $i$~\cite{PDG}.


Photon candidates are reconstructed following Ref.~\cite{Belle II:Ds2317}. Boosted-decision-tree classifiers are used to suppress misreconstructed photons from secondary hadronic interactions and beam-induced background~\cite{Cheema:2024iek}. For the signal channel $D_{s1}(2460)^{+} \to D_{s}^{*+}(\to D_{s}^{+}\gamma_{2})\gamma_{1}$, the energies of $\gamma_{1}$ and $\gamma_{2}$ are required to be greater than 0.50 and 0.14~GeV in the $e^{+}e^{-}$ c.m.\ frame, respectively. To suppress background from $\pi^{0}$ decays, the invariant mass of the $\gamma_{1}\gamma_{2}$ pair must satisfy $|M^{\rm rec}(\gamma_{1}\gamma_{2})-m(\pi^{0})|>15~{\rm MeV}/c^{2}$. The assignments of $\gamma_1$ and $\gamma_2$ can be clearly disambiguated, as the $D_s^+\gamma_1$ combination does not fall within the $D_s^{*+}$ signal region.
For the reference channel $D_{s1}(2460)^{+} \to D_{s}^{+}\gamma$, the energy of $\gamma$ is required to be greater than 0.60~GeV
in the $e^{+}e^{-}$ c.m.\ frame. For both decay channels, the invariant mass of any combination of a photon emitted directly from the $D_{s1}(2460)^{+}$ decay and other photon candidate in the rest of the event is required to lie outside a $15~{\rm MeV}/c^{2}$ window around the known $\pi^{0}$ mass~\cite{PDG}.
This requirement does not apply to the photon from the $D_s^{*+}$ decay, as it reduces the signal efficiency without improving the signal-to-background ratio.


The $D_{s}^{+}$ candidates are combined with a photon to form $D_{s}^{*+}$ candidates. To reduce the uncertainty from the $D_s^{+}$ mass 
resolution, we use the modified invariant mass $M(D_{s}^{+}\gamma_{2}) \equiv M^{\rm rec}(K^{+}K^{-}\pi^{+}\gamma_{2}) -M^{\rm rec}(K^{+}K^{-}\pi^{+}) + m(D_{s}^{+})$. 
This procedure improves the mass resolution by approximately 15\%, and a similar improvement is obtained for the reconstructed $D_{s1}(2460)^+$ candidates using the analogous modified invariant masses defined below.
The signal region for the $D_{s}^{*+}$ candidates is defined as $|M(D_{s}^{+}\gamma_{2})-m(D_{s}^{*+})|<15~{\rm MeV}/c^{2}$ (approximately 2.5 times the mass resolution), while the sidebands are defined as $|M(D_{s}^{+}\gamma_{2}) - m(D_{s}^{*+}) \pm 50| < 15~{\rm MeV}/c^{2}$, with a total width twice that of the signal region. 

The $D_{s}^{*+}\gamma_{1}$ and $D_{s}^{+}\gamma$ combinations are used to form $D_{s1}(2460)^{+}$ candidates for the signal and 
reference channels, respectively. A vertex-fitting algorithm is applied to the entire decay chain, with the requirement that the $D_{s1}(2460)^{+}$ 
candidate originates from the $e^{+}e^{-}$ interaction point~\cite{Belle-IIanalysissoftwareGroup:2019dlq}.
To reduce combinatorial backgrounds, the reduced momentum $x_{p}$ is required to be larger than 0.7, which also effectively suppresses $D_{s1}(2460)^{+}$
candidates from $B$ decays lying in the low-$x_p$ region. If multiple $D_{s1}(2460)^{+}$ candidates are found in an event, all combinations are retained for further analysis. 
The average candidate multiplicities for the signal and reference channels are $1.04$ ($1.05$) and $1.01$ ($1.01$) in Belle (Belle~II), respectively.
The corresponding multiplicities for all events in the data are consistent with those in the MC simulation.

\begin{figure}[htbp]
	\begin{center}
		\includegraphics[width=8.5cm]{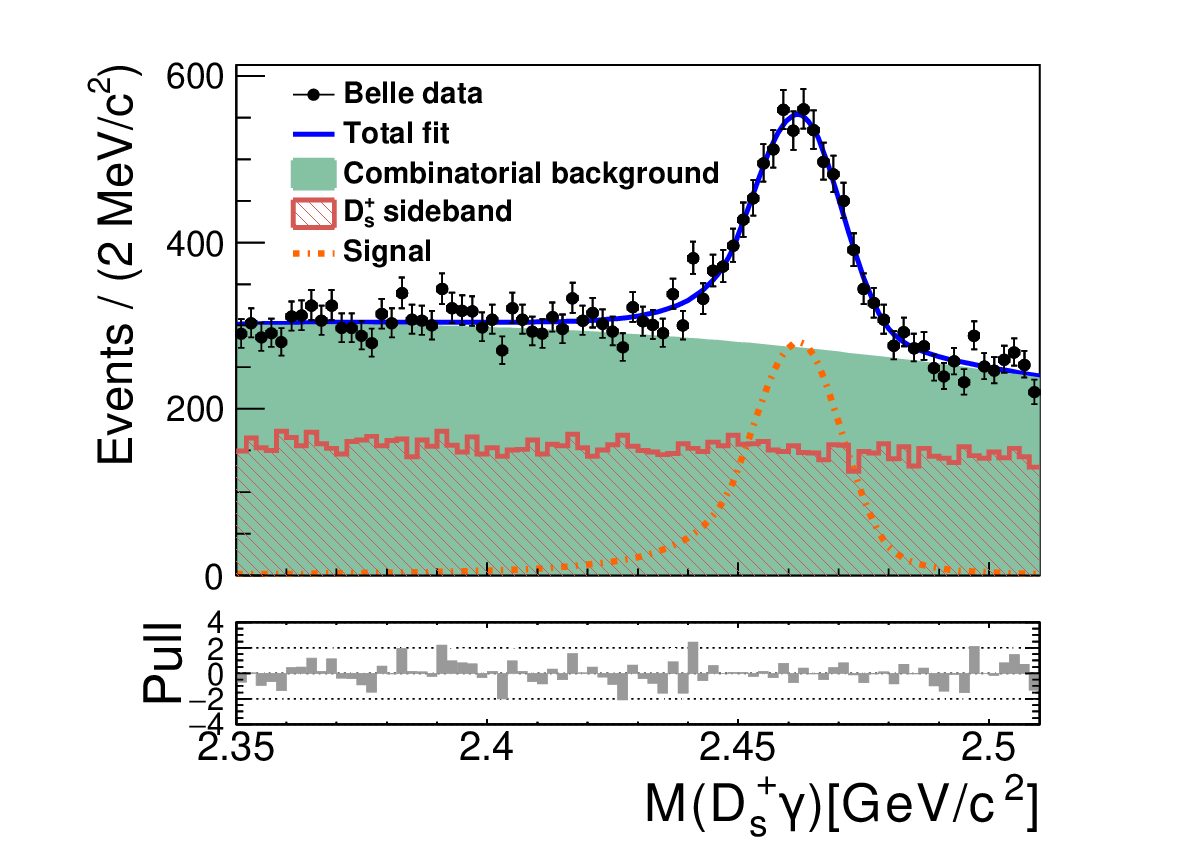}
		\put(-50,150){\bf (a)}
		  \put(-180,170){{\sffamily Belle Preliminary}\quad $\int \mathcal{L}\,\mathrm{d}t = 980~\mathrm{fb}^{-1}$}
		
		\vspace{0.5cm}
		\includegraphics[width=8.5cm]{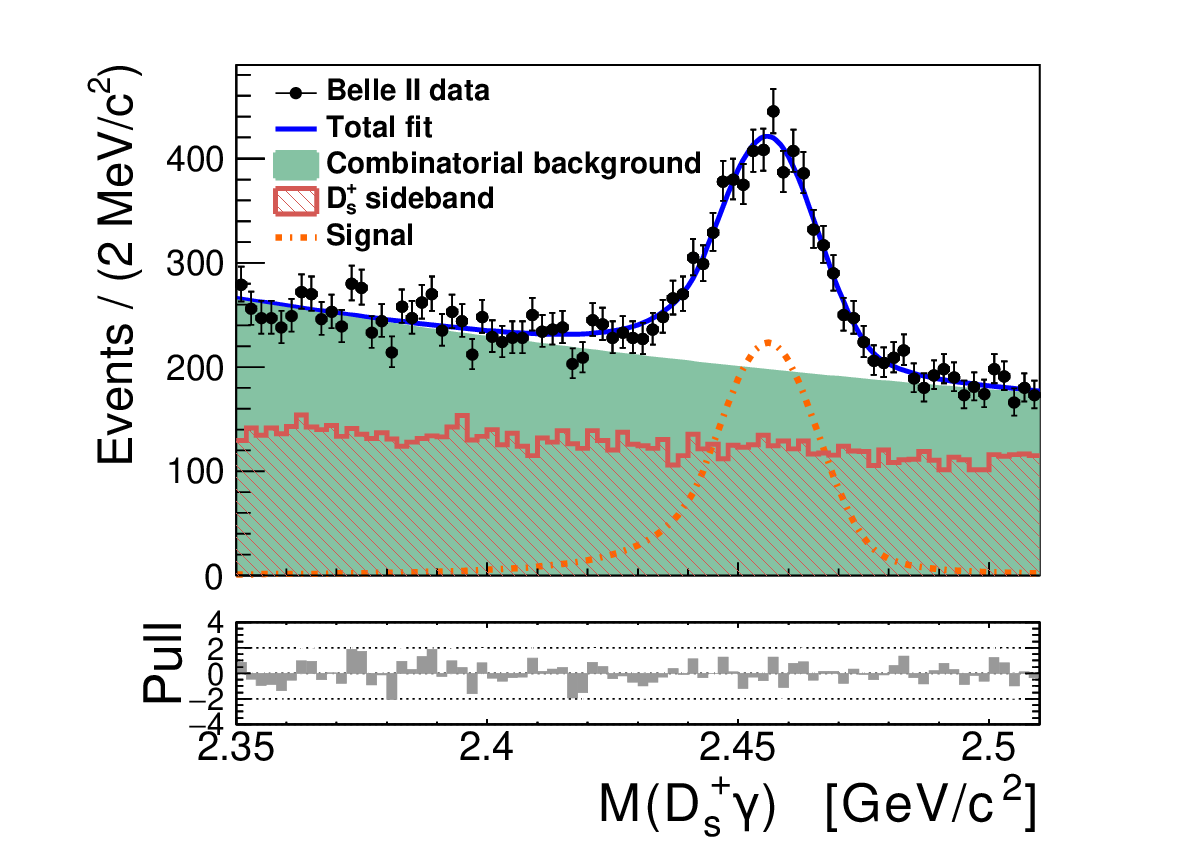}
		\put(-50,150){\bf (b)} 
		  \put(-180,170){{\sffamily Belle~II Preliminary}\quad $\int \mathcal{L}\,\mathrm{d}t = 572~\mathrm{fb}^{-1}$}
        \caption{Mass distributions of $M(D_{s}^{+}\gamma)$ for the reconstructed $D_{s1}(2460)^{+} \to D_{s}^{+}\gamma$ candidates with fit projections overlaid in (a) Belle and (b) Belle~II data. The data and fit components are indicated in the legends.
        The sideband distributions are scaled according to the ratio between the defined signal region and sideband of $D_{s}^{+}$.}\label{fig1}
	\end{center}
\end{figure}

After applying the aforementioned selection criteria, the $M(D_{s}^{+}\gamma)$ distributions for the reconstructed 
$D_{s1}(2460)^{+} \to D_{s}^{+}\gamma$ candidates in Belle and Belle~II data are shown in Fig.~\ref{fig1}, with the fit results overlaid. 
Here, $M(D_{s}^{+}\gamma)$ is defined as $M^{\rm rec}(K^{+}K^{-}\pi^{+}\gamma) - M^{\rm rec}(K^{+}K^{-}\pi^{+}) + m(D_{s}^{+})$, which reduces the uncertainty from the $D_{s}^{+}$ mass resolution.
Backgrounds are studied using both inclusive MC samples and data from the 
$D_{s}^{+}$ sidebands, and no peaking background is found in the $M(D_{s}^{+}\gamma)$ distributions. Clear $D_{s1}(2460)^{+} \to D_{s}^{+}\gamma$ signals are observed in Belle and Belle~II data samples. The signal yields are extracted separately for the two data samples using unbinned extended maximum-likelihood fits to the $M(D_s^+\gamma)$ distributions.
In each fit, a double-sided Crystal Ball (CB) function~\cite{Das:2016stf} is used as the signal probability density function (PDF), while the combinatorial background PDF is parametrized by a second-order Chebyshev polynomial.
All parameters of the signal and combinatorial background PDFs are allowed to float in the fit.
The pull distributions, defined as $(N^{\rm data}-N^{\rm fit})/\sqrt{N^{\rm data}}$, are also shown in Fig.~\ref{fig1}, where $N^{\rm data}$ is the number of events in each bin from data and $N^{\rm fit}$ is the number of events in each bin according to the fit. 

The fitted signal yields $N_{\rm exp}^{\rm fit}(D_{s}^{+}\gamma)$ are $(3.71 \pm 0.19)\times10^{3}$ and $(3.21 \pm 0.18)\times10^{3}$ for Belle and Belle~II, respectively, and the corresponding detection efficiencies $\varepsilon_{\rm exp}(D_{s}^{+}\gamma)$ are 10.1\% and 13.5\%. Here and in the following, the subscript ``exp'' denotes the experiment, Belle or Belle~II. Similar fits are performed to the $M(D_{s}^{+}\gamma)$ distributions in different $x_{p}$ bins to measure the $x_{p}$ distribution of the $D_{s1}(2460)^{+}$ candidate. The obtained efficiency-corrected $x_{p}$ distribution is used to derive an $x_{p}$-dependent correction to the signal MC simulation.

\begin{figure}[htbp]
	\begin{center}
		\includegraphics[width=8.5cm]{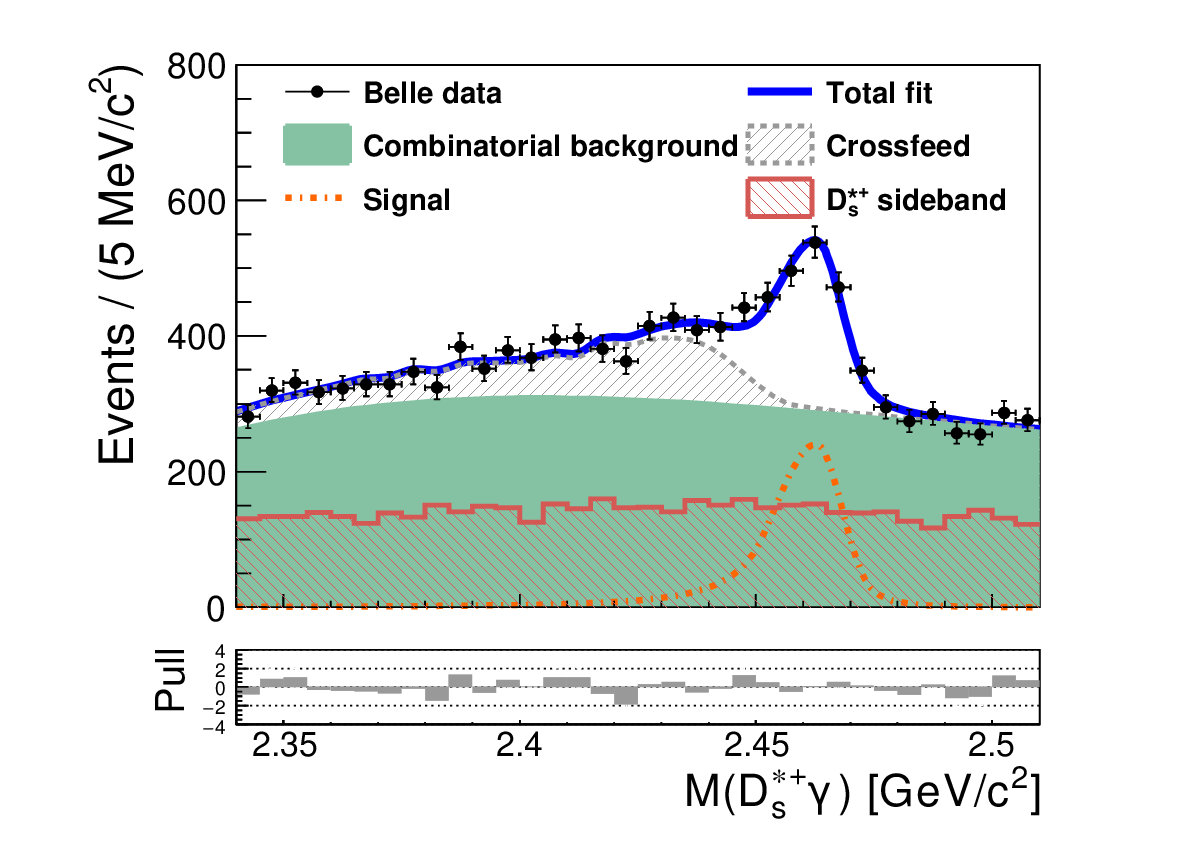}
		\put(-50,115){\bf (a)}
	    \put(-180,170){{\sffamily Belle Preliminary}\quad $\int \mathcal{L}\,\mathrm{d}t = 980~\mathrm{fb}^{-1}$}
		
		\includegraphics[width=8.5cm]{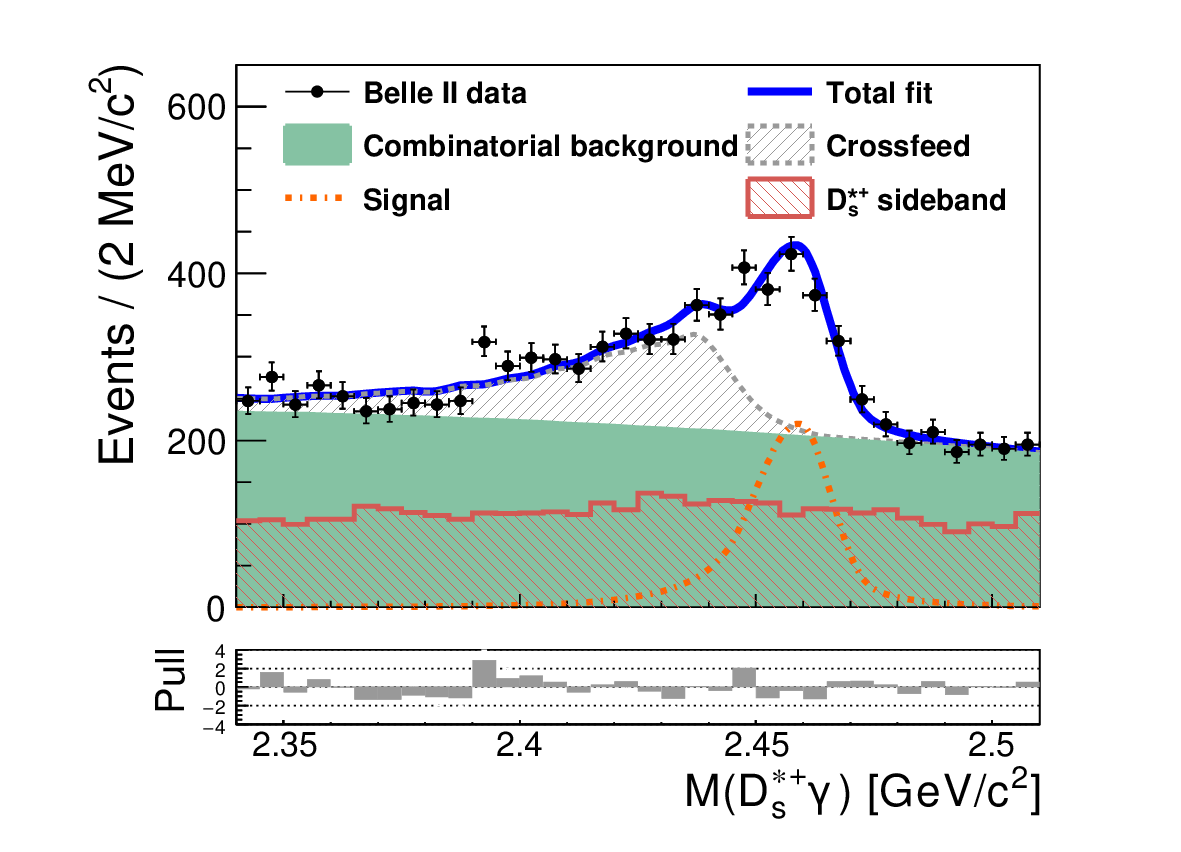}
		\put(-50,115){\bf (b)} 
        \put(-180,170){{\sffamily Belle~II Preliminary}\quad $\int \mathcal{L}\,\mathrm{d}t = 572~\mathrm{fb}^{-1}$}
        \caption{Mass distributions of $M(D_{s}^{*+}\gamma)$ for the reconstructed $D_{s1}(2460)^{+} \to D_{s}^{*+}\gamma$ candidates with fit projections overlaid in (a) Belle and (b) Belle~II data. The data and fit components are indicated in the legends. The sideband distributions are scaled according to the ratio between the defined signal region and sideband of $D_{s}^{*+}$.}\label{fig2}
	\end{center}
\end{figure}

Figure~\ref{fig2} shows the $M(D_{s}^{*+}\gamma)$ distributions for the reconstructed $D_{s1}(2460)^{+} \to D_{s}^{*+}\gamma$ candidates in Belle and Belle~II data. Here, $M(D_{s}^{*+}\gamma)$ is defined as $M^{\rm rec}(D_{s}^{+}\gamma_{1}\gamma_{2}) - M^{\rm rec}(D_{s}^{+}\gamma_{2}) + m(D_{s}^{*+})$, which reduces the uncertainty from the $D_{s}^{*+}$ mass resolution.
Clear $D_{s1}(2460)^{+}\to D_{s}^{*+}\gamma$ signal peaks are observed in both data samples, together with a broad enhancement on the low-mass side. Studies of inclusive MC samples show that the enhancement is due to crossfeed from $D_{s1}(2460)^{+}\to D_{s}^{*+}\pi^{0}$ decay in which one photon from $\pi^{0}\to\gamma\gamma$ is not reconstructed, and that the remaining background is dominated by a smooth combinatorial component. No peaking background is observed in the $D_{s}^{*+}$ sidebands in either the Belle or Belle~II data. Dedicated MC studies also show that other known decay channels of the $D_{s0}^{*}(2317)^{+}$ or $D_{s1}(2460)^{+}$ do not produce peaking backgrounds in the considered $M(D_s^{*+}\gamma)$ region.

To extract the branching-fraction ratio $\mathcal{R}$, we perform a simultaneous unbinned extended maximum-likelihood fit to the
$M(D_{s}^{*+}\gamma)$ distributions from Belle and Belle~II data, 
using the $D_{s1}(2460)^{+}\to D_s^{+}\gamma$ signal yields and the corresponding reconstruction efficiencies discussed earlier as inputs.
The $D_{s1}(2460)^{+} \to D_{s}^{*+}\gamma$ signal PDF is described by a double-sided CB function, while the combinatorial background PDF is parametrized by a third-order Chebyshev polynomial with free parameters. The crossfeed contribution from partially reconstructed $D_{s1}(2460)^{+}\to D_s^{*+}\pi^0$ decays is modeled using smoothed MC histograms, with the yields treated as free parameters.
For each data sample, the parameters of the signal PDF, except for the mean, are fixed to the values obtained from the corresponding MC simulation.
The value of $\mathcal{R}$ is shared as a common free parameter in the simultaneous fit, while the $D_{s1}(2460)^{+}\to D_{s}^{*+}\gamma$ signal yields are set separately for Belle and Belle~II according to $N_{\rm exp}(D_{s}^{*+}\gamma)=\mathcal{R}N_{\rm exp}^{\rm fit}(D_{s}^{+}\gamma) \varepsilon_{\rm exp}(D_{s}^{*+}\gamma)/\varepsilon_{\rm exp}(D_{s}^{+}\gamma)$.
Here, $\varepsilon_{\rm exp}(D_{s}^{*+}\gamma)$ is the detection efficiency of the $D_{s1}(2460)^{+}\to D_{s}^{*+} \gamma$ decay, with the branching fraction of $D_{s}^{*+} \to D_{s}^{+}\gamma$ included, which is $3.70\%$ for Belle and $5.42\%$ for Belle~II.
The fit results are shown in Fig.~\ref{fig2} along with the pull distributions. 
The fitted $\mathcal{R}$ value is $0.74 \pm 0.05$ with the statistical uncertainties on $N_{\rm exp}^{\rm fit}(D_{s}^{+}\gamma)$ 
included, and the corresponding signal yields of $D_{s1}(2460)^{+}\to D_{s}^{*+}\gamma$ in Belle and Belle~II data are
$(1.00 \pm 0.06)\times10^{3}$ and $(0.96 \pm 0.06)\times10^{3}$, respectively. The quoted uncertainties are statistical only.
 
The statistical significance of the $D_{s1}(2460)^{+}\to D_{s}^{*+}\gamma$ signal is $16.8\sigma$, 
estimated from the negative log-likelihood ratio $-2\ln(\mathcal{L}_{0}/\mathcal{L}_{\rm max}) = 292$~\cite{Wilks:1938dza},
while accounting for a difference of three degrees of freedom between the two fits.
Here, $\mathcal{L}_{0}$ and $\mathcal{L}_{\rm max}$ are the likelihood values obtained from fits without and with the signal component, respectively. After incorporating the systematic uncertainties discussed below, the signal significance remains greater than $10\sigma$.

As a consistency check, separate fits are performed to the Belle and Belle II data samples using the same fit model as in the simultaneous fit. The fitted signal yields $N_{\rm exp}^{\rm fit}(D_{s}^{*+}\gamma)$ are $(1.06 \pm 0.09)\times10^{3}$ and $(0.91 \pm 0.08)\times10^{3}$ for Belle and Belle~II, respectively. The corresponding $\mathcal{R}$ values are $0.79 \pm 0.08$ and $0.71 \pm 0.07$, respectively, which are consistent with each other. 
The quoted uncertainties are statistical only.
The efficiency-corrected yields per unit luminosity of the crossfeed component in Belle and Belle~II data are consistent within $1\sigma$. 
To further verify our findings, we fully reconstruct the decay $D_{s1}(2460)^{+} \to D_s^{*+}\pi^0$ using Belle data. Comparing the fitted signal yields and the relative reconstruction efficiencies for both the fully reconstructed and crossfeed samples, we find that the crossfeed yield agrees with the expectation within $1\sigma$.

The dominant systematic uncertainties in the measurement of $\mathcal{R}$ arise from the fit model, the photon reconstruction efficiency, 
the $x_{p}$ reweighting, and the potential $D_{s1}(2460)^+$ polarization.  The systematic uncertainties associated with the $D_{s}^{+}$ selection cancel in the extraction of $\mathcal{R}$. 

To estimate the systematic uncertainty associated with the fit model for the reference channel $D_{s1}(2460)^{+} \to D_{s}^{+}\gamma$, 
we repeat the fit with alternative configurations in which the fit range and the order of the background polynomial are varied by $\pm 10$ MeV and $\pm 1$ order, respectively. Based on the fitted signal and background components from each fit alternative, 1000 pseudo-experiments are generated and fitted with the nominal model. 
The largest difference between the mean fitted signal yield and the corresponding input yield is assigned as the systematic uncertainty on $N_{\rm exp}^{\rm fit}(D_{s}^{+}\gamma)$. 
The resulting fractional uncertainties are 1.4\% and 2.4\% for Belle and Belle~II, respectively. The reference channel yields are then varied by these amounts in the simultaneous fit to the $M(D_s^{*+}\gamma)$ distributions in data, resulting in a 1.9\% systematic uncertainty on $\mathcal{R}$.

The systematic uncertainty associated with the background shape for the signal channel $D_{s1}(2460)^{+}\to D_{s}^{*+}\gamma$ is evaluated using a procedure similar to that for the reference channel, yielding a 6.2\% uncertainty on $\mathcal{R}$. The uncertainty due to the fit range is estimated by extending the fit range to $[2.31, 2.60]$~{GeV}$/c^2$, while keeping the possible $D_{s1}(2536)^+$ signal region blinded. The resulting change in $\mathcal{R}$, $2\%$, is taken as the systematic uncertainty.
The data--MC difference in the $D_{s1}(2460)^+$ signal resolution is evaluated using the reference channel. For each experiment, the $M(D_s^+\gamma)$ distribution in data is fitted using the signal shape obtained from the corresponding signal MC sample, convolved with a Gaussian function whose width is allowed to vary. The fitted Gaussian widths are $2.2$ and $2.8~\mathrm{MeV}/{c^2}$ for Belle and Belle~II, respectively. For the signal channel, the simultaneous fit to the Belle and Belle~II data is repeated with the signal PDF convolved with a Gaussian function whose width is fixed to the value determined from the reference channel for each experiment, resulting in a 1.1\% systematic uncertainty on $\mathcal{R}$. 
The crossfeed shape is varied by changing the smoothing of the MC histogram, shifting its shape according to the data--MC difference in the $D_{s1}(2460)^{+}$ signal peak position, and convolving it with an additional Gaussian resolution function whose width is fixed to the data--MC difference in the mass resolution determined from the reference channel. The resulting changes in $\mathcal{R}$ are negligible.

The photon reconstruction efficiency is calibrated using radiative Bhabha events for Belle and radiative muon-pair and $\tau\to \pi/K \nu_\tau\pi^0$ control samples for Belle~II. 
The systematic uncertainties in the photon reconstruction efficiency are expected to be linearly correlated between the signal and reference channels, and therefore cancel in the overlapping photon momentum and polar-angle regions.
The residual systematic uncertainties from the non-overlapping regions on the signal-to-reference efficiency ratio, $\varepsilon_{\rm exp}(D_s^{*+}\gamma)/\varepsilon_{\rm exp}(D_s^{+}\gamma)$, are 4.2\% and 3.8\% for Belle and Belle~II, respectively. The efficiency ratios are varied within these uncertainties in the simultaneous fit to data. The resulting variation in $\mathcal{R}$ is 4.1\%.


The systematic uncertainty due to the $x_p$ reweighting is evaluated by varying the polynomial order used to fit the efficiency-corrected $x_p$ distribution by $\pm1$ around the nominal fourth-order choice. The signal MC samples are reweighted accordingly, and the resulting efficiencies are used in the simultaneous fit. The resulting variation in $\mathcal{R}$ is 0.9\%, which is assigned as the corresponding systematic uncertainty.

The systematic uncertainty associated with the potential $D_{s1}(2460)^+$ polarization is evaluated using the efficiency-corrected $\cos\theta_h$ distributions in data, where $\theta_h$ is the angle between the momentum of the $e^+e^-$ system and that of the photon in the $D_{s1}(2460)^+$ rest frame. The signal MC events for the reference and signal channels are reweighted using the efficiency-corrected $\cos\theta_h$ distributions obtained from data for the corresponding channels, and the resulting efficiencies are used in the simultaneous fit. The resulting variation in $\mathcal{R}$ is 0.9\%.

Assuming all systematic uncertainties are independent, they are added in quadrature, resulting in a total systematic uncertainty of $8.1\%$.

In summary, using data samples corresponding to integrated luminosities of 980 and 572~fb$^{-1}$ collected with the Belle and Belle~II detectors, respectively, we report the first observation of the radiative decay $D_{s1}(2460)^{+} \to D_s^{*+}\gamma$. 
Taking the decay $D_{s1}(2460)^{+} \to D_s^{+}\gamma$ as the reference channel, 
the branching-fraction ratio $\mathcal{R}=\mathcal{B}[D_{s1}(2460)^{+} \to D_{s}^{*+} \gamma]/\mathcal{B}[D_{s1}(2460)^{+} \to D_{s}^{+} \gamma]$
is measured to be $0.74 \pm 0.05 ({\rm stat.}) \pm 0.06 ({\rm syst.)}$.
Using the branching fraction $\mathcal{B}[D_{s1}(2460)^{+}\to D_s^{+}\gamma]=(18\pm4)\%$~\cite{PDG}, we obtain the absolute branching fraction $\mathcal{B}[D_{s1}(2460)^{+}\to D_s^{*+}\gamma] = [13\pm 1\,(\mathrm{stat.})\pm 1\,(\mathrm{syst.})\pm 3\,(\mathrm{ext.})]\%$, where the last uncertainty is due to the uncertainty in $\mathcal{B}[D_{s1}(2460)^{+}\to D_s^{+}\gamma]$. 
The measured branching fraction is approximately $2.8\sigma$ 
above the upper limit of 8\% established by the Particle Data Group at 90\% confidence level, after accounting for the correlated normalization uncertainty. Similarly, the measured $\mathcal{R}$ value differs from the CLEO~\cite{CLEO:2003ggt} and BaBar~\cite{BaBar:2006eep} measurements by $2.3\sigma$ and $2.1\sigma$, respectively.

\begin{figure}[htbp]
	\hspace{-0.5cm}
	\begin{overpic}[width=9.5cm]{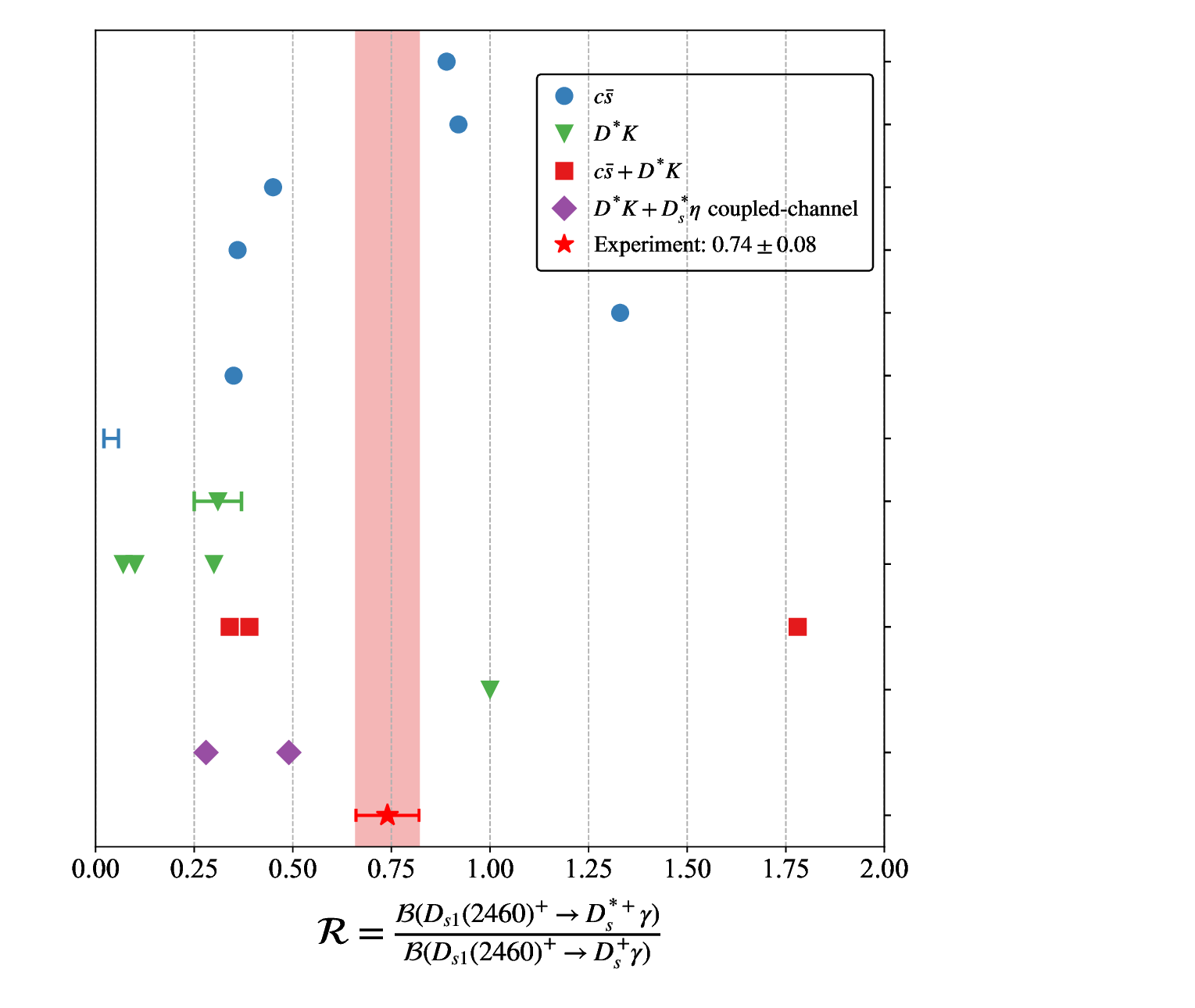} 	
    \put(77,78.00){\small Godfrey~\cite{Godfrey:2003kg}}		
    \put(77,72.75){\small Bardeen~\cite{Bardeen:2003kt}}
    \put(77,67.50){\small Ferrandes~\cite{Ferrandes:2004aa}}
    \put(77,62.25){\small Green~\cite{Green:2016occ}}	
    \put(77,57.00){\small Radford~\cite{Radford:2009bs}}
    \put(77,51.75){\small Bondar~\cite{Bondar:2025qzg}}
    \put(77,46.50){\small Colangelo~\cite{Colangelo:2005hv}}
    \put(77,41.25){\small Fu~\cite{Fu:2021wde}}
    \put(77,36.00){\small Zhang~\cite{Zhang:2024usz}}
    \put(77,30.75){\small Zhang~\cite{Zhang:2024usz}}
    \put(77,25.50){\small Close~\cite{Close:2005se}}
    \put(77,20.25){\small Lutz~\cite{Lutz:2007sk}}
    \put(77,15.00){\small Experiment}
	\end{overpic}

    \caption{Comparison of the measured $\mathcal{R}$ value with various theoretical predictions. For a conventional $c\bar{s}$ interpretation of the $D_{s1}(2460)^{+}$, the predictions are obtained from quark models by
    Godfrey~\cite{Godfrey:2003kg}, Green~\cite{Green:2016occ}, and
    Radford~\cite{Radford:2009bs};
    a heavy-light chiral effective theory by
    Bardeen~\cite{Bardeen:2003kt};
    heavy-quark effective theory combined with chiral perturbation theory by
    Ferrandes~\cite{Ferrandes:2004aa};
    a treatment of $^1\mathrm{P}_1$--$^3\mathrm{P}_1$ mixing by
    Bondar~\cite{Bondar:2025qzg};
    and light-cone QCD sum rules by
    Colangelo~\cite{Colangelo:2005hv}, with the shown $\mathcal{R}$ range inferred from the quoted radiative-width ranges.
    Predictions beyond the conventional $c\bar{s}$ picture include a $D^{*}K$ molecular description within effective field theory by Fu~\cite{Fu:2021wde};
    $D^{*}K$ molecular and mixed $c\bar{s}$--$D^{*}K$ configurations in an unquenched quark model by Zhang~\cite{Zhang:2024usz};
    a $D^*K$ molecular scenario considered by Close~\cite{Close:2005se};
    and a dynamically generated $D^{*}K$--$D_s^{*}\eta$ state within a chiral coupled-channel framework by Lutz~\cite{Lutz:2007sk}.
    For each of the two configurations considered by Zhang~\cite{Zhang:2024usz},
    the three predicted values correspond to the cutoff parameter
    $\Lambda_{0}=0.5$, $1.0$, and $1.5~\mathrm{GeV}$.
    For Lutz~\cite{Lutz:2007sk}, the two predicted values arise from the two possible
    sign choices of the electromagnetic coupling $e_{A}$ governing
    light-vector-meson radiative transitions.}\label{fig3}
\end{figure}

Figure~\ref{fig3} compares the measured $\mathcal{R}$ value with various theoretical predictions for different descriptions of the 
$D_{s1}(2460)^{+}$ state~\cite{Godfrey:2003kg, Bardeen:2003kt, Radford:2009bs, Ferrandes:2004aa, Green:2016occ, Bondar:2025qzg, Fu:2021wde, Colangelo:2005hv, Close:2005se, Zhang:2024usz, Lutz:2007sk}. 
The predictions show strong model dependence, spanning a wide range even for the same assumed internal structure. Our measurement provides new constraints on these
theoretical descriptions and new experimental input for understanding 
the internal structure of the $D_{s1}(2460)^{+}$ meson. 
Within a conventional $c\bar{s}$ interpretation, the measured $\mathcal{R}$ value also constrains the $^1\mathrm{P}_1$--$^3\mathrm{P}_1$ mixing in the $D_{s1}(2460)^{+}$ state. For example, in the quark-model framework from Ref.~\cite{Close:2005se}, it corresponds to a mixing angle of
$|\theta_s| = [48.4^{+0.9}_{-1.0}\,(\mathrm{stat.}) {}^{+1.1}_{-1.2}\,(\mathrm{syst.})]^\circ$, where only the experimental uncertainties are included.

This work, based on data collected using the Belle II detector, which was built and commissioned prior to March 2019, and data collected using the Belle detector, which was operated until June 2010,
was supported by
Higher Education and Science Committee of the Republic of Armenia Grant No.~23LCG-1C011;
Australian Research Council and Research Grants
No.~DP200101792, 
No.~DP210101900, 
No.~DP210102831, 
No.~DE220100462, 
No.~LE210100098, 
and
No.~LE230100085; 
Austrian Federal Ministry of Education, Science and Research,
Austrian Science Fund (FWF) Grants
DOI:~10.55776/P34529,
DOI:~10.55776/J4731,
DOI:~10.55776/J4625,
DOI:~10.55776/M3153,
and
DOI:~10.55776/PAT1836324,
and
Horizon 2020 ERC Starting Grant No.~947006 ``InterLeptons'';
Natural Sciences and Engineering Research Council of Canada, Digital Research Alliance of Canada, and Canada Foundation for Innovation;
National Key R\&D Program of China under Contract No.~2024YFA1610503,
and
No.~2024YFA1610504
National Natural Science Foundation of China and Research Grants
No.~11575017,
No.~11761141009,
No.~11705209,
No.~11975076,
No.~12135005,
No.~12150004,
No.~12161141008,
No.~12405099,
No.~12475093,
and
No.~12175041,
and Shandong Provincial Natural Science Foundation Project~ZR2022JQ02;
the Czech Science Foundation Grant No. 22-18469S,  Regional funds of EU/MEYS: OPJAK
FORTE CZ.02.01.01/00/22\_008/0004632 
and
Charles University Grant Agency project No. 246122;
European Research Council, Seventh Framework PIEF-GA-2013-622527,
Horizon 2020 ERC-Advanced Grants No.~267104 and No.~884719,
Horizon 2020 ERC-Consolidator Grant No.~819127,
Horizon 2020 Marie Sklodowska-Curie Grant Agreement No.~700525 ``NIOBE''
and
No.~101026516,
and
Horizon Europe Marie Sklodowska-Curie Staff Exchange project JENNIFER3 Grant Agreement No.~101183137 (European grants);
L’Institut National de Physique Nucl\'eaire et de Physique des
Particules (IN2P3) du CNRS under Project Identification No.
CNRS-IN2P3-14-PP-033
and L’Agence Nationale de la Recherche (ANR) under Grant No. ANR-23-CE31-
0018 and ANR-25-CE31-1333 (France);
BMFTR, DFG, HGF, MPG, and AvH Foundation (Germany);
Department of Atomic Energy under Project Identification No.~RTI 4002,
Department of Science and Technology,
and
UPES SEED funding programs
No.~UPES/R\&D-SEED-INFRA/17052023/01 and
No.~UPES/R\&D-SOE/20062022/06 (India);
Israel Science Foundation Grant No.~2476/17,
U.S.-Israel Binational Science Foundation Grant No.~2016113, and
Israel Ministry of Science Grant No.~3-16543;
Istituto Nazionale di Fisica Nucleare and the Research Grants BELLE2,
and
the ICSC – Centro Nazionale di Ricerca in High Performance Computing, Big Data and Quantum Computing, funded by European Union – NextGenerationEU;
Japan Society for the Promotion of Science, Grant-in-Aid for Scientific Research Grants
No.~16H03993,
No.~16H06492,
No.~16K05323,
No.~17H01133,
No.~17H05405,
No.~18K03621,
No.~18H03710,
No.~18H05226,
No.~19H00682, 
No.~20H05850,
No.~20H05858,
No.~22H00144,
No.~22K14056,
No.~22K21347,
No.~23H05433,
No.~26220706,
No.~26400255,
and
No.~26H02056,
and
the Ministry of Education, Culture, Sports, Science, and Technology (MEXT) of Japan;  
National Research Foundation (NRF) of Korea Grants
No.~2021R1-F1A-1064008,
No.~2022R1-A2C-1003993,
No.~RS-2018-NR031074,
No.~RS-2021-NR060129,
No.~RS-2024-00354342,
No.~RS-2025-02219521,
No.~RS-2026-25471491,
No.~RS-2026-25480677,
and
No.~RS-2026-25486791,
Radiation Science Research Institute,
Foreign Large-Size Research Facility Application Supporting project,
the Global Science Experimental Data Hub Center, the Korea Institute of Science and
Technology Information (K26L1M2C3)
and
KREONET/GLORIAD;
Universiti Malaya RU grant, Akademi Sains Malaysia, and Ministry of Education Malaysia;
Frontiers of Science Program Contracts
No.~FOINS-296,
No.~CB-221329,
No.~CB-236394,
No.~CB-254409,
and
No.~CB-180023, and SEP-CINVESTAV Research Grant No.~237 (Mexico);
the Polish Ministry of Science and Higher Education and the National Science Center;
the Ministry of Science and Higher Education of the Russian Federation
and
the HSE University Basic Research Program, Moscow;
University of Tabuk Research Grants
No.~S-0256-1438 and No.~S-0280-1439 (Saudi Arabia);
Slovenian Research Agency and Research Grants
No.~J1-50010
and
No.~P1-0135;
Basque Government through grant IT1977-26,Ikerbasque, Basque Foundation for Science,
State Agency for Research of the Spanish Ministry of Science, Innovation and Universities (MICIU/AEI) through
Grants No. PID2022-136510NB-C33, PID2025-174203NB-I00, and the Severo Ochoa project CEX2023-001292-S,
State Secretariat for  Telecommunications and Digital Infrastructure with reference TSI-069100-2023-0012, Spain;
The Knut and Alice Wallenberg Foundation (Sweden), Contracts No.~2021.0174, No.~2021.0299, and No.~2023.0315;
National Science and Technology Council,
and
Ministry of Education (Taiwan);
Thailand Center of Excellence in Physics;
TUBITAK ULAKBIM (Turkey);
National Research Foundation of Ukraine, Project No.~2020.02/0257,
and
Ministry of Education and Science of Ukraine;
the U.S. National Science Foundation and Research Grants
No.~PHY-1913789 
and
No.~PHY-2111604, 
and the U.S. Department of Energy and Research Awards
No.~DE-AC06-76RLO1830, 
No.~DE-SC0007983, 
No.~DE-SC0009824, 
No.~DE-SC0009973, 
No.~DE-SC0010007, 
No.~DE-SC0010073, 
No.~DE-SC0010118, 
No.~DE-SC0010504, 
No.~DE-SC0011784, 
No.~DE-SC0012704, 
No.~DE-SC0019230, 
No.~DE-SC0026613, 
No.~DE-SC0026215, 
No.~DE-SC0022350, 
No.~DE-SC0023470; 
and
the Vietnam Academy of Science and Technology (VAST) under Grant
No.~DL0000.05/26-27.

These acknowledgements are not to be interpreted as an endorsement of any statement made
by any of our institutes, funding agencies, governments, or their representatives.

We thank the SuperKEKB team for delivering high-luminosity collisions;
the KEK cryogenics group for the efficient operation of the detector solenoid magnet and IBBelle on site;
the KEK Computing Research Center for on-site computing support; the NII for SINET6 network support;
and the raw-data centers hosted by BNL, DESY, GridKa, IN2P3, INFN, 
and the University of Victoria.

\textit{Data availability} --- \ifdefvoid{\hepdata}{}{Numerical data corresponding to the results presented are available as HEPData.}
The full Belle II data are not publicly available. The collaboration will consider requests for access to the data that support this article.

\end{document}